\documentclass[reqno,a4paper,final,11pt]{amsart}
\usepackage[utf8x]{inputenc}
\usepackage{amssymb,amstext,amsthm,amscd,eucal}
\usepackage{color}
\usepackage{a4wide}
\usepackage{array}
\usepackage[all]{xy}
\usepackage{hyperref}
\hypersetup{%
  pdftitle   = {Deformations of M-theory Killing superalgebras},
  pdfkeywords = {M-theory, supergravity, supersymmetry, superalgebra,
    deformations}
  pdfauthor  = {José Figueroa-O'Farrill},
  pdfcreator = {\LaTeX\ with package \flqq hyperref\frqq}
}
\PrerenderUnicode{éÉ}
\newcommand{\half}{\tfrac12}
\newcommand{\fg}{\mathfrak{g}}
\newcommand{\fM}{\mathfrak{M}}

\newcommand{\fh}{\mathfrak{h}}
\newcommand{\fs}{\mathfrak{s}}
\newcommand{\fz}{\mathfrak{z}}
\newcommand{\fso}{\mathfrak{so}}
\newcommand{\fco}{\mathfrak{co}}
\newcommand{\fosp}{\mathfrak{osp}}
\newcommand{\fsu}{\mathfrak{su}}
\newcommand{\fu}{\mathfrak{u}}
\newcommand{\Cl}{\mathrm{C}\ell}
\newcommand{\Spin}{\mathrm{Spin}}

\ifx\Sp\UnDeFiNeD
\newcommand{\Sp}{\mathrm{Sp}}
\else
\renewcommand{\Sp}{\mathrm{Sp}}
\fi

\newcommand{\U}{\mathrm{U}}
\newcommand{\RR}{\mathbb{R}}

\newcommand{\ZZ}{\mathbb{Z}}

\renewcommand{\1}{\boldsymbol{1}}

\newcommand{\be}{\boldsymbol{e}}
\newcommand{\beps}{\boldsymbol{\varepsilon}}
\newcommand{\bnu}{\boldsymbol{\nu}}
\newcommand{\bv}{\boldsymbol{v}}
\newcommand{\bw}{\boldsymbol{w}}

\DeclareMathOperator{\AdS}{AdS}

\DeclareMathOperator{\End}{End}

\newcommand{\MUNCH}[1]{\relax}

\allowdisplaybreaks[1]
\begin{document}
\title[Deformations of Killing superalgebras]{Deformations of M-theory
  Killing superalgebras}
\author{José Figueroa-O'Farrill}
\address{Maxwell Institute and School of Mathematics, University of
  Edinburgh, UK}
\email{J.M.Figueroa@ed.ac.uk}
\thanks{EMPG-07-11}
\begin{abstract}
  We classify the Lie superalgebra deformations of the Killing
  superalgebras of some M-theory backgrounds.  We show that the
  Killing superalgebras of the Minkowski, Freund--Rubin and M5-brane
  backgrounds are rigid, whereas the ones for the M-wave, the
  Kaluza--Klein monopole and the M2-brane admit deformations, which we
  give explicitly.
\end{abstract}
\maketitle
\tableofcontents

\section{Introduction}

The main article of faith justifying much of the present research on
supergravity is that supergravity may teach us something about string
theory.  In particular, it is assumed that supergravity backgrounds
may be corrected to yield exact string backgrounds, something which
could perhaps be proved, at least in special cases, using techniques
from the study of partial differential equations such as the Banach or
Nash--Moser implicit function theorems.  We will not question this
assumption in this paper.  Instead we would like to explore how
invariants of a supergravity background can change as the background
itself gets deformed to incorporate the quantum corrections.  We will
focus on one such invariant: the \emph{Killing superalgebra} of the
background \cite{AFHS, GMT1, GMT2, PKT, JMFKilling, FOPFlux, NewIIB,
  ALOKilling, FMPHom, EHJGMHom}, a Lie superalgebra so called because
it is constructed out of Killing vectors and Killing spinors.

For supergravities which arise as limits of M- or string theories, it
is a natural question to ask what happens to the Killing superalgebra
under stringy (i.e., $\alpha'$) or M-theoretic quantum corrections.
There seems to be some evidence \cite{GPTHeterotic, LPSCYcorr,
  LPSTG2corr, LPSTholcorr, Hyakutake1, LPSghcorr, Hyakutake2,
  Hyakutake3} supporting the persistence of the notion of Killing
superalgebra under this procedure; although to be fair the study of
quantum corrections is still very much in its infancy and there is not
enough data to argue this point convincingly.  Let us however assume
that the notion persists in some way.  Then surely one should find the
Killing superalgebra of a quantum-corrected background among the
\emph{deformations} (in the sense of Gerstenhaber \cite{Gerstenhaber})
of the Killing superalgebra of the classical background or, allowing
for symmetry breaking, of a suitable subsuperalgebra.  It remains to
decide what algebraic structure one should deform.

A Lie superalgebra can be viewed in many equivalent ways.  It is
standard to view it as a vector superspace with a skewsymmetric
bracket obeying the Jacobi identity, but by going to the
universal enveloping algebra we can also view it as an
associative algebra or more generally as a cocommutative Hopf
algebra.  Conversely every cocommutative Hopf algebra generated
by its primitive elements is the universal enveloping algebra of
a Lie superalgebra.  Dually, we may also view a Lie superalgebra
structure on a vector space $V$ as a differential graded superalgebra
structure on $\Lambda^\bullet V^*$, whose differential has degree $1$.
Conversely every such differential is dual to a Lie superalgebra
structure on $V$.  The question is then how to deform a Lie
superalgebra: as a Lie superalgebra? as a Hopf algebra? or as a
differential graded superalgebra?  In the first case we remain in the
world of Lie superalgebras, whereas the other two cases would bring us
to the worlds of quantum supergroups and $L_\infty$ superalgebras,
respectively.  From our present position of ignorance, the safest
assumption and, in any case, the one we will explore in this paper, is
to remain within the category of Lie superalgebras.

Therefore in this paper we will classify the possible \emph{Lie
  superalgebra deformations} of the Killing superalgebras of some
M-theory backgrounds: all maximally supersymmetric backgrounds except
for the Kowalski-Glikman wave, and the elementary half-BPS
backgrounds: M2- and M5-branes, as well as the M-wave and the
Kaluza--Klein monopole.  The calculations employ established
homological techniques which we will briefly review below.

These calculations may also be of use in classical supergravity.
Indeed, deformation is an inverse process to that of contraction; that
is, the deformations of a Lie superalgebra $\fg$ consist of all the
Lie superalgebras which contract to $\fg$ analytically.  We know that
under certain geometric limits, such as the plane-wave limit
\cite{PenrosePlaneWave,GuevenPlaneWave}, the Killing superalgebra of a
background gets contracted \cite{ShortLimits,Limits,HatKamiSaka,FSPL}.
Hence classifying the possible deformations of the Killing
superalgebra of a background gives us hints about the existence of
other nearby backgrounds of which the background in question can be a
geometric limit.  Of course, reconstructing the background from its
Killing superalgebra is only ever possible if the dimension of the
superalgebra is large enough to constrain the geometry sufficiently.
Research is in progress \cite{FigRitDef} to investigate the existence
of classical M-theory backgrounds whose Killing superalgebras are the
deformations found in this paper.

This paper is organised as follows.  In Section~\ref{sec:deformcohom}
we will discuss the basics of Lie superalgebra cohomology and the
basic technique to compute the possible deformations, based on the
Hochschild--Serre spectral sequence.  In Section~\ref{sec:minkowski}
we prove that the Killing superalgebra of the Minkowski background is
rigid, in contrast with the four-dimensional situation.  Appealing to
general results, we deduce in Section~\ref{sec:FR}, that the
Freund--Rubin superalgebras too are rigid.  In
Section~\ref{sec:branes} we explore the Lie superalgebra deformations
of the Killing superalgebras for the elementary M2- and M5-brane.  We
find that whereas the Killing superalgebra of the M5-brane is rigid,
that of the M2-brane admits an integrable one-parameter deformation
suggesting that the worldvolume of the membrane deforms to $\AdS_3$.
In Section~\ref{sec:KSAgrav} we do the same for the Killing
superalgebras of the elementary half-BPS purely gravitational
backgrounds: the M-wave and the Kaluza--Klein monopole, and find that
whereas the M-wave superalgebra admits an integrable one-parameter
deformation, the Kaluza--Klein monopole superalgebra admits two such
families: one is reminiscent of a nongeometric background, whereas
the other suggests that the Minkowski factor deforms to $\AdS_7$.
Finally in Section~\ref{sec:conclusion} we offer some concluding
remarks.

\section{Lie superalgebra deformations and cohomology}
\label{sec:deformcohom}

\subsection{Deformations}
\label{sec:deform}

Recall that a Lie superalgebra is a vector superspace $\fg =
\fg_0 \oplus \fg_1$, together with an even bilinear map $[-,-]:\fg
\times \fg \to \fg$ which is \emph{skewsymmetric}
\begin{equation}
  \label{eq:skew}
  [X,Y]= - (-1)^{XY} [Y,X]~,
\end{equation}
and satisfies the \emph{Jacobi identity}
\begin{equation}
  \label{eq:jacobi}
  [X,[Y,Z]] = [[X,Y],Z] + (-1)^{XY}[Y,[X,Z]]~,
\end{equation}
for homogeneous $X,Y,Z\in\fg$ and where in the expression for signs we
denote the grading of a homogeneous $X\in\fg$ also by $X$.

By a \textbf{Lie superalgebra deformation} of $\fg$, we mean a
one-parameter family of Lie superalgebra structures $[-,-]_t$ on $\fg$
depending analytically on $t$ and agreeing at $t=0$ with the original
Lie superalgebra structure $[-,-]$.  Expanding the bracket $[-,-]_t$
in a power series in $t$ we find
\begin{equation}
  [X,Y]_t = [X,Y] + t \Phi_1(X,Y) + t^2 \Phi_2(X,Y) + \cdots =
  \sum_{n\geq 0} t^n \Phi_n(X,Y)~,
\end{equation}
with $\Phi_0(X,Y) = [X,Y]$.  The skewsymmetry condition
(\ref{eq:skew}) says that
\begin{equation}
  \Phi_k(X,Y) = -(-1)^{XY} \Phi_k(Y,X)
\end{equation}
for all $k$, whereas the Jacobi identity gives rise to an infinite
number of equations, one for each power of $t$:
\begin{equation}
  \label{eq:deformeqs}
    \sum_{\ell+m=n} \left( \Phi_\ell(X,\Phi_m(Y,Z)) -
      \Phi_\ell(\Phi_m(X,Y),Z) - (-1)^{XY} \Phi_\ell(Y,\Phi_m(X,Z))
    \right) = 0~,
\end{equation}
for all $n\geq 0$.  The first equation, for $n=0$, is the Jacobi
identity for $\Phi_0 = [-,-]$ and for $n>0$ we obtain equations for
the higher $\Phi_k$.  In particular, the equation
\begin{multline}
  \label{eq:2cocycle}
  \Phi_1(X,[Y,Z]) - \Phi_1([X,Y],Z) - (-1)^{XY} \Phi_1(Y,[X,Z])\\
  + [X,\Phi_1(Y,Z)] - [\Phi_1(X,Y),Z] - (-1)^{XY} [Y,\Phi_1(X,Z)] =
  0~,
\end{multline}
for $n=1$ is a condition on $\Phi_1: \Lambda^2 \fg \to \fg$, which can
be interpreted as a cocycle condition in the cochain complex
$C^2(\fg;\fg)$ to be defined below.  A $\Phi_1$ obeying equation
(\ref{eq:2cocycle}) is said to be an \emph{infinitesimal deformation}.
Such an infinitesimal deformation is said to be \emph{trivial}, if it
is the result of a $t$-dependent change of basis; in other words, if
$\Phi_1$ is given by
\begin{equation}
  \label{eq:2coboundary}
  \Phi_1(X,Y) = [X,B(Y)] - (-1)^{XY} [Y,B(X)] - B([X,Y])~,
\end{equation}
for some even linear transformation $B:\fg \to \fg$.  It is easy to
check that such $\Phi_1$ automatically obeys (\ref{eq:2cocycle}).
Indeed, equation (\ref{eq:2coboundary}) says that $\Phi_1$ is a
coboundary in $C^2(\fg;\fg)$.  The space of (nontrivial) infinitesimal
deformations is therefore the space of solutions $\Phi_1$ of
(\ref{eq:2cocycle}) factored by the space of $\Phi_1$ given by
(\ref{eq:2coboundary}), which can be reinterpreted as the cohomology
group $H^2(\fg;\fg)$ to be defined below.  The further equations in
(\ref{eq:deformeqs}) for higher $n$ give obstructions to integrating
the infinitesimal deformation.  They can be reinterpreted as a
sequence of cohomology classes in $H^3(\fg;\fg)$.  In a nutshell, the
tangent space to the moduli space of deformations of a Lie
superalgebra $\fg$ is given by $H^2(\fg;\fg)$, whereas the
obstructions to integrating a deformation along a direction in
$H^2(\fg;\fg)$ are given by a sequence of classes in $H^3(\fg;\fg)$
and which are in the image of a squaring map $H^2(\fg;\fg) \to
H^3(\fg;\fg)$ described in \cite{NijenhuisRichardson}.  For example,
the equation for $n=2$ says that the $3$-cocycle
\begin{equation}
  \label{eq:obstruction}
  [\Phi_1,\Phi_1](X,Y,Z) := \Phi_1(X,\Phi_1(Y,Z)) -
      \Phi_1(\Phi_1(X,Y),Z) - (-1)^{XY} \Phi_1(Y,\Phi_1(X,Z))
\end{equation}
obtained by ``squaring'' $\Phi_1$ should be a coboundary (of
$\Phi_2$), et~cetera.

\subsection{Cohomology}
\label{sec:cohom}

Lie superalgebra cohomology was introduced by Le{\u\i}tes
\cite{Leites} and is reviewed in \cite[§1.6]{FuksCohomology}.  It is a
straight-forward extension of the better-known Lie algebra cohomology
theory of Chevalley and Eilenberg \cite{ChevalleyEilenberg}.

Let $\fg = \fg_0 \oplus \fg_1$ be a finite-dimensional real Lie
superalgebra and let $\fM = \fM_0 \oplus \fM_1$ be a $\fg$-module.  We
will let $X\cdot m$ denote the action of $X \in \fg$ on $m \in \fM$.
We demand that the action preserve the parity, so that $\fg_\alpha
\cdot \fM_\beta \subset \fM_{\alpha + \beta}$.  Let $C^n(\fg;\fM)$
denote the space of multilinear maps
\begin{equation}
 f : \underbrace{\fg \times \cdots \times \fg}_{n~\text{times}} \to \fM
\end{equation}
satisfying the following skewsymmetry condition:
\begin{equation}
  f(X_1,\dots,X_n) = - (-1)^{X_iX_{i+1}}
  f(X_1,\dots,X_{i-1},X_{i+1},X_i,X_{i+2},\dots,X_n)~.
\end{equation}
The vector space $C^n(\fg;\fM)$ of such maps is naturally
$\ZZ_2$-graded.  We will let
\begin{equation}
  C(\fg;\fM) = \bigoplus_{n=0}^\infty C^n(\fg;\fM)~.
\end{equation}
If $\fh < \fg$ is an ideal, then each vector space $C^n(\fh;\fM)$ is
naturally a $\fg$-module, where for all $Y \in \fg$, $f \in
C^n(\fh;\fM)$ and $X_1,\dots,X_n \in \fh$,
\begin{multline}
  (Y \cdot f)(X_1,\dots,X_n) = Y \cdot f(X_1,\dots,X_n) \\
  - \sum_{i=1}^n (-1)^{Y(f+X_1 + \cdots + X_{i-1})}
  f(X_1,\dots,[Y,X_i],\dots,X_n)~.
\end{multline}
We define the differential $d: C^n(\fg;\fM) \to C^{n+1}(\fg;\fM)$ as
follows.  If $m \in C^0(\fg;\fM) = \fM$,
\begin{equation}
  \label{eq:dm}
  (d m)(X) = (-1)^{Xm} X \cdot m~,
\end{equation}
and if $f \in C^{n>0}(\fg;\fM)$,
\begin{multline}
  (d f)(X_0,X_1,\dots,X_n)
  = \sum_{i=0}^n (-1)^{i+X_i(f+X_0 + \cdots +
    X_{i-1})} X_i \cdot f(X_0,\dots,\widehat{X_i},\dots,X_n) \\
  + \sum_{0\leq i<j \leq n} (-1)^{i+j + (X_i+X_j)(X_0 + \dots +
    X_{i-1}) + X_j(X_{i+1} + \cdots + X_{j-1})} \\
  \times f([X_i,X_j],X_0,\dots, \widehat{X_i}, \dots, \widehat{X_j},
  \dots, X_n)~.
\end{multline}
Notice that $d$ has zero parity.  It obeys $d^2 = 0$ and it is
$\fg$-equivariant, so that $X \cdot df = d ( X \cdot f)$ for all $X
\in \fg$ and $f\in C^n(\fg;\fM)$.  For every $X \in \fg$ we define a
derivation $\imath_X : C^n(\fg;\fM) \to C^{n-1}(\fg;\fM)$ by
\begin{equation}
  \label{eq:iX}
  (\imath_X f)(X_1,\dots,X_{n-1}) = (-1)^{fX} f(X,X_1,\dots,X_{n-1})~.
\end{equation}
It follows easily that
\begin{equation}
  \imath_X (Y \cdot f) - (-1)^{XY} Y \cdot \imath_X f = \imath_{[X,Y]}f
\end{equation}
and also that the Cartan formula holds
\begin{equation}
  \imath_X df + d\imath_X f = X \cdot f~.
\end{equation}

Let $(X_a,X_i)$ and $(m_A,m_I)$ denote homogeneous bases for $\fg =
\fg_0 \oplus \fg_1$ and $\fM = \fM_0 \oplus \fM_1$, respectively.
Here and in what follows we will adhere to the summation convention.
Doing so, we have
\begin{equation}
  \begin{aligned}[m]
    X_a \cdot m_A &= K_{aA}^B m_B\\
    X_a \cdot m_I &= K_{aI}^J m_J
  \end{aligned}
\qquad\text{and}\qquad
  \begin{aligned}[m]
    X_i \cdot m_A &= K_{iA}^I m_I\\
    X_i \cdot m_I &= K_{iI}^A m_A~,
  \end{aligned}
\end{equation}
and also
\begin{equation}
  [X_a,X_b] = f_{ab}^c X_c~,\quad
  [X_a,X_i] = f_{ai}^j X_j \quad\text{and}\quad
  [X_i,X_j] = f_{ij}^a X_a~.
\end{equation}
Let $(\theta^a,\theta^i)$ denote the canonical dual basis for $\fg^* =
\fg_0^* \oplus \fg_1^*$.  The following rules, together with the fact
that $d$ is a derivation, suffice to compute the differential on any
cochain in $C^n(\fg;\fM)$:
\begin{equation}
  \begin{aligned}[m]
    d\theta^a &= - \half f_{bc}^a \theta^b \wedge \theta^c
    + \half  f_{ij}^a \theta^i \wedge \theta^j\\
    d\theta^i &= - f_{aj}^i \theta^a \wedge \theta^j
  \end{aligned}
  \qquad\qquad
  \begin{aligned}[m]
    dm_A &= \theta^a \otimes K_{aA}^B m_B - \theta^i \otimes K_{iA}^I
    m_I\\
    dm_I &= \theta^a \otimes K_{aI}^J m_J - \theta^i \otimes K_{iI}^A
    m_A~.
  \end{aligned}
\end{equation}
Notice that our convention for $\wedge$ is that $\alpha \wedge \beta = -
(-1)^{\alpha\beta} \beta \wedge \alpha$, so that it is
superskewsymmetric; in particular, $\theta^i \wedge \theta^j = \theta^j
\wedge \theta^i$.  As a check of these formulae, it may be shown that the
differential of the identity map $\fg \to \fg$, thought of as the
$1$-cochain $\theta^a\otimes X_a - \theta^i\otimes X_i  \in
C^1(\fg;\fg)$, is the $2$-cochain
\begin{equation}
  \half f_{ab}{}^c \theta^a \wedge \theta^b \otimes X_c - f_{ai}{}^j
  \theta^a \wedge \theta^i \otimes X_j - \half f_{ij}{}^a \theta^i
  \wedge \theta^j \otimes X_a \in C^2(\fg;\fg)
\end{equation}
corresponding to the Lie bracket.

\subsection{Hochschild--Serre factorisation theorem}
\label{sec:HS}

A fundamental tool in computing these cohomology groups is the
Hochschild--Serre spectral sequence \cite{HochschildSerre} which
exploits the existence of a semisimple factor $\fs$ of $\fg$ in order
to reduce the calculation of the cohomology to that of the much
smaller subcomplex of $\fs$-invariants.  This method was used in
\cite{JMFGalilean} to calculate the possible Lie algebra deformations
of the Galilean algebras.  The superalgebra version of this theorem is
discussed in \cite[§1.6.5]{FuksCohomology}; although it has also
appeared in \cite{Binegar, TripathyPatra}.  In \cite{Binegar} it was
used in order to compute the possible deformations of the
four-dimensional Poincaré superalgebra and $\fosp(4|2)$; although the
deformed Poincaré algebra in that paper is actually incorrect.  A
correct calculation of the unique deformation \cite{ZuminoAdS} of the
four-dimensional Poincaré superalgebra appears in
\cite{TripathyPatra}, which also contains a fuller treatment of the
Hochschild--Serre spectral sequence.  In a nutshell, the theorem
allows us to work covariantly with respect to any semisimple
subalgebra of the Lie superalgebra in question.

More precisely, let $\fg$ be a finite-dimensional real Lie
superalgebra and let $\fM$ denote a $\fg$-module.  Let $I < \fg$ be an
ideal such that $\fs := \fg/I$ is a semisimple Lie \emph{algebra}.
Then the factorisation theorem of Hochschild--Serre states that
\begin{equation}
  H^n(\fg;\fM) \cong \bigoplus_{i=0}^n\left( H^{n-i}(\fs) \otimes
  H^i(I;\fM)^{\fs}\right)~,
\end{equation}
where $H^\bullet(I;\fM)^{\fs}$ is the cohomology of the subcomplex
$C^\bullet(I;\fM)^{\fs}$ of $\fs$-invariant cochains and
$H^\bullet(\fs)$ is the cohomology with values in the trivial
one-dimensional module.  Using the Whitehead lemma, $H^j(\fs) = 0$ for
$j=1,2$, and the fact that $H^0(\fs) \cong \RR$, the above direct sum
simplifies to
\begin{equation}
  H^n(\fg;\fM) \cong H^n(I;\fM)^{\fs} \oplus
  \bigoplus_{i=0}^{n-3}\left( H^{n-i}(\fs) \otimes
    H^i(I;\fM)^{\fs}\right)~.
\end{equation}
In particular, we have that
\begin{equation}
  H^1(\fg;\fg) \cong H^1(I;\fg)^{\fs} \qquad\text{and}\qquad
  H^2(\fg;\fg) \cong H^2(I;\fg)^{\fs}~,
\end{equation}
whereas
\begin{equation}
  H^3(\fg;\fg) \cong H^3(I;\fg)^{\fs} \oplus H^3(\fs)\otimes \fz~,
\end{equation}
where $\fz=\fg^\fg$ is the centre of $\fg$.  
Of course, the full strength of this theorem is only ever felt if
$\fg$ admits a sufficiently large semisimple factor $\fs$.

\section{Rigidity of the Poincaré superalgebra}
\label{sec:minkowski}

As a first calculation, let us take $\fg$ to be the $11$-dimensional
Poincaré superalgebra, which is the Killing superalgebra of the
Minkowski background of $11$-dimensional supergravity.  We take $I$ to
be the supertranslation ideal consisting of the momentum generators
and the supercharges.  The semisimple factor is the Lorentz subalgebra
$\fs \cong \fso(10,1)$.  As $\fs$-modules, $I = V \oplus \Delta$ and
$\fg = V \oplus \Lambda^2 V \oplus \Delta$, where $V$ is the
11-dimensional vector representation and $\Delta$ is the real
$32$-dimensional irreducible representation of spinors.

It will prove convenient \emph{not} to identify $I$ and $V \oplus
\Delta$.  We will let $P$, $Q$ denote the isomorphisms between $V$ and
$\Delta$ and the corresponding subspaces of $I$, and similarly we will
let $L:\Lambda^2 V \to \fso(V)$ denote the natural isomorphism.

Let $\be_a$ be a basis for $V$ and $\beps_i$ be a basis for $\Delta$.
We will choose an action of the Clifford algebra $\Cl(V)$ on $\Delta$
once and for all.  Following the time-honoured tradition, the image of
$\be_a$ under the embedding $V \to \Cl(V)$ will be denoted
$\gamma_a$.  Our conventions are $\gamma_a \gamma_b + \gamma_b
\gamma_a = + 2 \eta_{ab} \1$, with $\eta_{ab}$ mostly plus.  More
traditionally still, we will let $\gamma_{ab\dots c}$ denote the image
of $\be_a \wedge \be_b \wedge \cdots \wedge \be_c$ under the vector
space isomorphism $\Lambda V \stackrel{\cong}{\longrightarrow}
\Cl(V)$.

The corresponding bases for $I$ are $P_a := P(\be_a)$ and $Q_i =
Q(\beps_i)$.  We will let $\be^a$ be the canonical dual basis of
$V^*$.  Because $\fs$ leaves invariant the Minkowski metric $\eta \in
S^2V^*$, we may identify $V$ and $V^*$ by ``raising/lowering indices''
with $\eta$ and our notation reflects this.  Similarly we let
$\beps^i$ denote the canonical dual basis for $\Delta^*$, where we may
again identify $\Delta$ and $\Delta^*$ using the $\fs$-invariant
symplectic form on $\Delta$.  Letting $P$ and $Q$ also stand for the
isomorphisms of $V^*$ and $\Delta^*$ with the corresponding subspaces
of $I^*$, we will let $P^a = P(\be^a)$ and $Q^i = Q(\be^i)$.  Finally
we will also let $L_{ab}:=L(\be_a \wedge \be_b)$, for $a<b$,  define a
basis for $\fs$.

The Poincaré superalgebra consists of a Lorentz subalgebra spanned by
the $L_{ab}$ and in addition
\begin{equation}
  \begin{aligned}[m]
    [L_{ab},Q_i] &= \half \gamma_{ab} \cdot Q_i\\
    [L_{ab},P_c] &=  \eta_{bc} P_a - \eta_{ac} P_b\\
    [Q_i, Q_j] &= \gamma^a_{ij} P_a~,
  \end{aligned}
\end{equation}
where
\begin{equation}
  \gamma_{ab} \cdot Q_i = Q(\gamma_{ab}\cdot \beps_i) = Q_j
  (\gamma_{ab})^j{}_i~,
\end{equation}
and similarly for the action of any other element in the Clifford
algebra $\Cl(V)$, and
\begin{equation}
  \gamma^a_{ij} := \left<\beps_i, \gamma^a \cdot \beps_j\right>~,
\end{equation}
where $\left<-,-\right>$ is the $\fs$-invariant symplectic structure
on $\Delta$.

Let us investigate the subcomplex $C^\bullet :=
C^\bullet(I;\fg)^{\fs}$ of Lorentz-invariant cochains in
$C^\bullet(I;\fg)$ in low dimension.  For applications to the theory
of Lie superalgebra deformations we are interested only in \emph{even}
cochains; however this is not a restriction because of the
representation-theoretic ``spin statistics'' theorem, which states
that there are no Lorentz-invariant maps between ``fermionic'' and
``bosonic'' representations.

There are no Lorentz-invariant $0$-cochains, since $\fg$ contains no
Lorentz scalars.  The even cochains in $C^1(I;\fg)$ are maps $V \to V
\oplus \Lambda^2 V$ and $\Delta \to \Delta$.  Since $V$ and $\Delta$
are irreducible representations, Schur's lemma says that the only
equivariant maps are multiples of the identity maps $V \to V$ and
$\Delta \to \Delta$.  Therefore, a basis for $C^1$ is given by the
vectors
\begin{equation}
  P^a \otimes P_a \qquad\text{and}\qquad Q^i \otimes Q_i~.
\end{equation}

The even $2$-cochains are maps of the form $\Lambda^2 V \to V \oplus
\Lambda^2V$, $V \otimes \Delta \to \Delta$ and $S^2\Delta \to V \oplus
\Lambda^2V$.  Lorentz invariance is again very restrictive and
there is a four-dimensional subspace of equivariant maps which is
spanned by the identity map $\Lambda^2 V \to \Lambda^2V$, Clifford
multiplication $V \otimes \Delta \to \Delta$ and the spinor squaring
maps $S^2 \Delta \to V$ and $S^2\Delta \to \Lambda^2V$.  A basis for 
$C^2$ is given by
\begin{equation}
  \begin{aligned}[m]
  & P^a \wedge P^b \otimes L_{ab}\\
  & P^a \wedge Q^i \otimes (\gamma_a \cdot Q_i)
  \end{aligned}
\qquad\qquad
  \begin{aligned}[m]
  &Q^i \wedge Q^j \otimes \gamma^a_{ij} P_a\\
  &Q^i \wedge Q^j \otimes \gamma^{ab}_{ij} L_{ab}~.
  \end{aligned}
\end{equation}

The even $3$-cochains are given by maps of the form $\Lambda^3 V \to V
\oplus \Lambda^2 V$, $\Lambda^2V \otimes \Delta \to \Delta$, $V
\otimes S^2\Delta \to V \oplus \Lambda^2 V$ and $S^3\Delta \to
\Delta$.  The Lorentz-equivariant maps are given in terms of the natural
operations.  There is no Lorentz-equivariant map $\Lambda^3V \to V
\oplus \Lambda^2 V$ because all three representations $V$, $\Lambda^2
V$ and $\Lambda^3 V$ are irreducible and non-equivalent.  The only
equivariant map $\Lambda^2 V \otimes \Delta \to \Delta$ is the spin
representation, for which a representative cochain is given by
\begin{equation}
  P^a \wedge P^b \wedge Q^i \otimes \gamma_{ab} \cdot Q_i~.
\end{equation}
The equivariant maps $V \otimes S^2\Delta \to V \oplus \Lambda^2 V$
are given by the compositions
\begin{equation}
  \begin{CD}
    V \otimes S^2\Delta @>>> V \otimes (V \oplus \Lambda^2 V) @>\wedge
    \oplus \imath>>
    \Lambda^2V \oplus V~,
  \end{CD}
\end{equation}
whose representative cochains are
\begin{equation}
  P^a \wedge Q^i \wedge Q^j \otimes \gamma^b_{ij} L_{ab}
  \quad\text{and}\quad
  P^a \wedge Q^i \wedge Q^j \otimes (\gamma_a{}^b)_{ij} P_b~.
  \end{equation}
Finally, the equivariant maps $S^3\Delta \to \Delta$ have the following
representative cochains:
\begin{equation}
  Q^i \wedge Q^j \wedge Q^k \otimes \gamma^a_{ij} \gamma_a \cdot Q_k
  \quad\text{and}\quad
  Q^i \wedge Q^j \wedge Q^k \otimes \gamma^{ab}_{ij} \gamma_{ab} \cdot Q_k~.  
\end{equation}
We see that $C^3$ is therefore five dimensional.

The differential $d$ in the invariant subcomplex
\begin{equation}
  \begin{CD}
    0 @>>> C^1 @>d>> C^2 @>d>> C^3 @>>> \cdots
  \end{CD}
\end{equation}
is defined by its action on the elements of $I^*$ and of $\fg$ as an
$I$-module; that is,
\begin{equation}
  \begin{aligned}[m]
    d P^a &= \half \gamma^a_{ij} Q^i \wedge Q^j\\
    d Q^i &= 0\\
    d P_a &= 0\\
    d Q_i &= - \gamma^a_{ij} Q^j \otimes  P_a\\
    d L_{ab} &= \eta_{ac} P^c \otimes P_b - \eta_{bc} P^c \otimes P_a + \half
    Q^i \otimes \gamma_{ab} \cdot Q_i~.
  \end{aligned}
\end{equation}

Acting on the $1$-cochains, we see that
\begin{equation}
  \begin{aligned}[m]
    d(P^a \otimes P_a) &= \half \gamma^a_{ij} Q^i \wedge Q^j \otimes  P_a\\
    d(Q^i \otimes Q_i) &= \gamma^a_{ij} Q^i \wedge Q^j \otimes P_a~.
  \end{aligned}
\end{equation}
Thus $2 P^a\otimes P_a - Q^i \otimes Q_i$ is a cocycle, whence, in
the absence of any coboundaries, we conclude that $H^1(\fg;\fg) \cong
\RR$.  This corresponds to an outer derivation of $\fg$ given by
dilatations.  The corresponding extension is obtained by replacing
$\fso(10,1)$ by $\fco(10,1)$ acting on $\fg$ in such that way that
$L_{ab}, Q_i, P_a$ have weights $0$, $1$ and $2$, respectively.

We also learn that the $2$-cochain $\gamma^a_{ij} Q^i \wedge Q^j
\otimes P_a$ is a coboundary.  The differential of the remaining three
$2$-cochains are
\begin{equation}
  \begin{aligned}[m]
    d(P^a \wedge P^b \otimes L_{ab}) &= \gamma^a_{ij} Q^i \wedge Q^j
    \wedge P^b \otimes L_{ab} + \half P^a \wedge P^b \wedge Q^i
    \otimes
    \gamma_{ab} \cdot Q_i~,\\
    d(P^a \wedge Q^i \otimes (\gamma_a \cdot Q_i)) &=
    (\gamma_a{}^b)_{ij} P^a \wedge Q^i \wedge Q^j \otimes P_b + \half
    \gamma^a_{ij} Q^i \wedge Q^j \wedge Q^k \otimes \gamma_a \cdot Q_k~,\\
    d(\gamma^{ab}_{ij} Q^i \wedge Q^j \otimes L_{ab}) &= 2
    (\gamma_a{}^b)_{ij} P^a \wedge Q^i \wedge Q^j \otimes P_b + \half
    \gamma^{ab}_{ij} Q^i \wedge Q^j \wedge Q^k \otimes \gamma_{ab}
    \cdot Q_k~.
  \end{aligned}
\end{equation}
The only possible cocycle would be
\begin{equation}
  \Psi := 2 P^a \wedge Q^i \otimes (\gamma_a \cdot Q_i) -
  \gamma^{ab}_{ij} Q^i \wedge Q^j \otimes L_{ab}~,
\end{equation}
whose differential is
\begin{equation}
  \label{eq:cocyclep}
  d\Psi = Q^i \wedge Q^j \wedge Q^k \otimes \left(\half
    \gamma^{ab}_{ij} \gamma_{ab} \cdot Q_k - \gamma^a_{ij} \gamma_a
    \cdot Q_k \right)~.
\end{equation}
By the usual polarisation identity, which says that if $P \in S^3V^*$
and $p(\bv) = P(\bv,\bv,\bv)$ is the associated cubic form, then
\begin{multline}
  P(\bv_1,\bv_2,\bv_3) = \tfrac16 \bigl( p(\bv_1 + \bv_2 + \bv_3) -
    p(\bv_1 + \bv_2) - p(\bv_1 + \bv_3)\\  - p(\bv_2 + \bv_3) + p(\bv_1)
    + p(\bv_2) + p(\bv_3) \bigr)~,
\end{multline}
equation \eqref{eq:cocyclep} would vanish if and only if for all
spinors $\psi \in \Delta$,
\begin{equation}
  \left<\psi, \gamma^a \cdot \psi\right> \gamma_a \cdot \psi - \half
  \left<\psi, \gamma^{ab} \cdot \psi\right> \gamma_{ab} \cdot \psi
  \stackrel{?}{=} 0~.
\end{equation}
This is a Lorentz-covariant equation, whence it is zero for a $\psi
\in \Delta$ it will be zero for every other spinor in its Lorentz
orbit.  There are two possible orbits (apart from the trivial orbit
consisting of the zero spinor, for which this equation is trivially
satisfied).  The above identity holds for the small orbit consisting
of spinors whose Dirac current is null, but it does not hold for the
generic orbit consisting of spinors whose Dirac current is timelike.
Indeed, for all $\psi \in \Delta$, one finds that
\begin{equation}
  \label{eq:fierz}
  \left<\psi, \gamma^a \cdot \psi\right> \gamma_a \cdot \psi +
  \tfrac1{10} \left<\psi, \gamma^{ab} \cdot \psi\right> \gamma_{ab}
  \cdot \psi = 0~;
\end{equation}
although for $\psi$ in the small orbit both terms vanish separately.
Therefore we conclude that $H^2(\fg;\fg) = 0$ and the $11$-dimensional
Poincaré superalgebra is rigid.

This is in sharp contrast with the four-dimensional case.  As shown in
\cite{Binegar,TripathyPatra}, the four-dimensional superalgebra admits
a deformation \cite{ZuminoAdS} whose bosonic subalgebra is the isometry
algebra of anti de~Sitter spacetime.

This result is consistent with the fact that the Minkowski vacuum does not
receive M-theoretic corrections, which follows from the observation
that corrections to the equations of motions come in the 
shape of polynomials of the curvature and the field strength, both of
which vanish for this background.

It is well-known that the Minkowski background arises as various
limits of the other maximally supersymmetric backgrounds.  These
limits are known to contract the Killing superalgebra, whence one
might expect to discover deformations of the Killing superalgebra of
the Minkowski background which reverse these contractions and hence
one might be puzzled by the rigidity found above.  The solution to the
puzzle is to notice that the dimension of the Killing superalgebras of
the Freund--Rubin \cite{FreundRubin,AdS7S4} or Kowalski-Glikman
\cite{KG} backgrounds is $(38|32)$, whereas that of the Minkowski
background is $(66|32)$.  There are subalgebras of the Poincaré
superalgebra, namely the image of the contractions of the Killing
superalgebras of the other maximally supersymmetric backgrounds, which
do admit deformations, but the full superalgebra does not.  This shows
that one must exercise care when concluding the existence or otherwise
of corrected supergravity backgrounds based solely on the existence of
deformations of the corresponding Killing superalgebras.

On the other hand, the Killing superalgebra of the Kowalski-Glikman
background does have deformations, which are the Killing superalgebras
of the Freund--Rubin backgrounds; although as we will now argue, the
superalgebras of the latter backgrounds are actually rigid.

\subsection{Rigidity of the Freund--Rubin superalgebras}
\label{sec:FR}

As reviewed for example in \cite{JMFKilling}, the Killing
superalgebras of the Freund--Rubin backgrounds $\AdS_4 \times S^7$ and
$\AdS_7 \times S^4$ are $\fosp(8|4)$ and $\fosp(6,2|4)$, respectively,
which are real forms of the complex Lie superalgebra of type $D(4,2)$
in the notation of Kac \cite{KacSuperSketch}, whose Killing form is
nondegenerate.  The proof of the rigidity of semisimple Lie algebras
uses the nondegeneracy of the Killing form to construct a chain
homotopy, whence we expect that for Lie superalgebras with
nondegenerate Killing form, the same result should obtain.  Indeed,
this has already been shown in \cite{SSSLjahovski}, from where we deduce
the rigidity of the Killing superalgebras of the Freund--Rubin
backgrounds.  This agrees with the fact that these backgrounds do not
receive quantum corrections \cite{KalRajVacua}.

\section{Superalgebra deformations of brane backgrounds}
\label{sec:branes}

In this section we detail the calculations of $H^2(\fg;\fg)$ for the
Killing superalgebras of the $\half$-BPS maximally symmetric M2- and
M5-brane backgrounds.  The Killing superalgebras are subsuperalgebras
of the one for the Minkowski background, to which the branes are
asymptotic.

\subsection{A deformation of the M2-brane Killing superalgebra}
\label{sec:M2}

The Killing superalgebra $\fg = \fg_0 \oplus \fg_1$ of the M2-brane is
the subalgebra of the Poincaré superalgebra defined as follows.  First
we split the $11$-dimensional lorentzian vector space $V = W \oplus
W^\perp$, where $W$ is three-dimensional lorentzian.  The subalgebra
of $\fso(V)$ which preserves this split is $\fso(W) \oplus
\fso(W^\perp)$.  Then $\fg_0 = \fso(W) \oplus \fso(W^\perp) \oplus W$.
The odd part of the superalgebra is the subspace $\Delta$ of the
$\fso(V)$ spinor module $\Delta(V)$ consisting of those spinors $\psi$
for which $\bnu_W \cdot \psi = \psi$, where $\bnu$ is the element in the
Clifford algebra representing the volume form.  As an $\fso(W) \oplus
\fso(W^\perp)$-module, this is $\Delta(W) \otimes \Delta(W^\perp)_-$,
where the chirality condition on the $\fso(W^\perp)$-spinor comes
about because $\fg_1$ is the subspace of the irreducible
$\Cl(V)$-module consisting of spinors $\psi$ for which $\bnu_V\cdot
\psi = - \psi$ and $\bnu_{W} \cdot \psi = \psi$, whence
$\bnu_{W^\perp} \cdot \psi = - \psi$.  The resulting Lie superalgebra
has dimension $(34|16)$.

In order to apply the Hochschild--Serre factorisation theorem, we will
choose $I$ to be the supertranslation ideal, which is isomorphic to $W
\oplus \Delta$, so that $\fs = \fso(W)\oplus \fso(W^\perp)$.  We will
let $\be_\mu$ and $\be_a$ be a basis for $W$ and $W^\perp$,
respectively, and let $\beps_i$ be a basis for $\Delta$.  Unlike in
the previous section, here $i$ only goes from $1$ to $16$.  We will
let $P_\mu$ and $Q_i$ denote the corresponding bases for $I$ and
$L_{\mu\nu}$ and $L_{ab}$ the corresponding bases for $\fs$.  As
before we will let $P^\mu$ and $Q^i$ denote the bases for $I^*$
canonically dual to $P_\mu$ and $Q_i$, respectively.

In this basis, the Lie brackets are
\begin{equation}
  \label{eq:SLAM2}
  \begin{aligned}[m]
    [L_{\mu\nu} , Q_i] &= \half \gamma_{\mu\nu}\cdot Q_i\\
    [L_{ab} , Q_i] &= \half \gamma_{ab}\cdot Q_i\\
    [L_{\mu\nu}, P_\rho] &= \eta_{\nu\rho} P_\mu - \eta_{\mu\rho}
    P_\nu\\
    [Q_i, Q_j] &= \gamma^\mu_{ij} P_\mu
  \end{aligned}
\end{equation}
in addition to the ones of $\fs$.

There are no $\fs$-invariant elements in $\fg$, whence there are no
invariant $0$-cochains.  The space $C^1$ of invariant $1$-cochains is
three-dimensional, spanned by the identity maps $W \to W$ and $\Delta
\to \Delta$, as well as the natural $\fs$-equivariant isomorphism $W
\to \Lambda^2 W$ induced from Hodge duality.  The representative
cochains are
\begin{equation}
  P^\mu \otimes P_\mu~, \quad Q^i \otimes Q_i \quad\text{and}\quad
  \varepsilon_\rho{}^{\mu\nu} P^\rho \otimes L_{\mu\nu}~.
\end{equation}

The space $C^2$ of invariant $2$-cochains is six-dimensional, spanned by
the Hodge duality map $\Lambda^2 W \to W$ and the identity map
$\Lambda^2W \to \Lambda^2W$, as well as Clifford product $W \otimes
\Delta \to \Delta$, and the three squaring maps $S^2\Delta  \to W
\oplus \Lambda^2W \oplus \Lambda^2W^\perp$.  The representative
cochains are
\begin{equation}
  \begin{aligned}[m]
    &\varepsilon_{\mu\nu}{}^\rho P^\mu \wedge P^\nu \otimes P_\rho\\
    &P^\mu \wedge P^\nu \otimes L_{\mu\nu}\\
    &P^\mu \wedge Q^i \otimes \gamma_\mu \cdot Q_i
  \end{aligned}
  \qquad\qquad
  \begin{aligned}[m]
    &\gamma^\mu_{ij} Q^i \wedge Q^j \otimes P_\mu\\
    &\gamma^{\mu\nu}_{ij} Q^i \wedge Q^j \otimes L_{\mu\nu}\\
    &\gamma^{ab}_{ij} Q^i \wedge Q^j \otimes L_{ab}~.
  \end{aligned}
\end{equation}
Notice the term $\varepsilon_{\mu\nu\rho} P^\mu \wedge Q^i \otimes
\gamma^{\nu\rho}\cdot Q_i$ is omitted, due to the fact that $\bnu_W
\cdot Q_i = Q_i$, whence $\gamma_\mu \cdot Q_i$ and
$\varepsilon_{\mu\nu\rho} \gamma^{\nu\rho} \cdot Q_i$ are
proportional.  Indeed, $\varepsilon_{\mu\nu\rho} \gamma^\rho \cdot Q_i
= - \gamma_{\mu\nu} \cdot Q_i$ and $\half \varepsilon_{\mu\nu\rho}
\gamma^{\nu\rho} \cdot Q_i = \gamma_\mu \cdot Q_i$.

The space $C^3$ of invariant $3$-cochains is four-dimensional.  It is
spanned by the following natural maps:
\begin{itemize}
\item $\Lambda^2 W \otimes \Delta \to \Delta$, given by the $\fso(W)$
  action;
\item $W \otimes S^2\Delta \to W \oplus \Lambda^2 W$, given by the
  squaring map on spinors;
\item $S^3\Delta \to \Delta$, given by the composition
  \begin{equation}
    \begin{CD}
      S^3 \Delta @>>> W \otimes \Delta @>>> \Delta~,
    \end{CD}
  \end{equation}
  where the first map is the squaring of the spinors and the second is
  made out of Clifford multiplication by $W$ on $\Delta$.
\end{itemize}

The absence of any $\Lambda^2 W^\perp$ in the above cochains should not
have gone unnoticed by the attentive reader.  It is not hard to show
that there is no equivariant map $W\otimes S^2\Delta \to
\Lambda^2 W^\perp$, since the corresponding cochain $\left<\beps_i,\gamma_\mu
  \cdot \gamma^{ab} \cdot \beps_j\right> P^\mu \wedge Q^i \wedge Q^j
\otimes L_{ab}$ vanishes because of the skew-symmetry (in $ij$) of
$\left<\beps_i,\gamma_\mu \cdot \gamma^{ab} \cdot \beps_j\right>$.  Similarly,
the composition $S^3 \Delta \to \Lambda^2 W^\perp \otimes \Delta \to
\Delta$ can be written as a linear combination of the composition
$S^3 \Delta \to W \otimes \Delta \to \Delta$, by virtue of
\eqref{eq:fierz}.  Indeed, unpacking \eqref{eq:fierz} under $\fso(W)
\oplus \fso(W^\perp)$, we find
\begin{multline}
  \left<\psi, \gamma^\mu \cdot \psi\right> \gamma_\mu \cdot \psi 
  + \left<\psi, \gamma^a \cdot \psi\right> \gamma_a \cdot \psi 
  + \tfrac1{10} \left<\psi, \gamma^{ab} \cdot \psi\right> \gamma_{ab}
  \cdot \psi \\
  + \tfrac1{10} \left<\psi, \gamma^{\mu\nu} \cdot \psi\right> \gamma_{\mu\nu}
  \cdot \psi 
  + \tfrac1{5} \left<\psi, \gamma^a \cdot \gamma^\mu \cdot \psi\right>
  \gamma_a \cdot \gamma_\mu \cdot \psi = 0~.
\end{multline}
The condition $\bnu_W \cdot \psi = \psi$ says that
\begin{equation}
\left<\psi, \gamma^{\mu\nu} \cdot \psi\right> \gamma_{\mu\nu}
  \cdot \psi  = - 2 \left<\psi, \gamma^\mu \cdot \psi\right> \gamma_\mu
  \cdot \psi~,
\end{equation}
This shows why we did not list the composition $S^3 \Delta \to
\Lambda^2 W \otimes \Delta \to \Delta$ among the maps above.

In addition, the self-adjointness of $\bnu_W$ relative to the spinor
inner product and the relations $\bnu_W \cdot \bw = \bw 
\cdot \bnu_W$ for $\bw \in W$ and $\bnu_W \cdot \bv = - \bv \cdot
\bnu_W$ for $\bv \in W^\perp$, means that $\left<\psi, \gamma^a \cdot
  \psi\right> = 0$ and $\left<\psi,   \gamma^a \cdot \gamma^\mu \cdot
  \psi\right>=0$, whence the identity \eqref{eq:fierz} can be
rewritten as
\begin{equation}
  \left<\psi, \gamma^{ab} \cdot \psi\right> \gamma_{ab} \cdot \psi  =
  -8 \left<\psi, \gamma^\mu \cdot \psi\right> \gamma_\mu \cdot \psi~.
\end{equation}
By the usual polarisation trick, this rules out the existence of the
extra cochains involving $\Lambda^2 W^\perp$.

An explicit basis for the invariant $3$-cochains is given by
\begin{equation}
  \begin{aligned}[t]
    &P^\mu \wedge P^\nu \wedge Q^i \otimes \gamma_{\mu\nu} \cdot Q_i\\
    & \gamma^\mu_{ij} Q^i \wedge Q^j \wedge Q^k \otimes \gamma_\mu
    \cdot Q_k
  \end{aligned}
  \qquad\qquad
  \begin{aligned}[t]
    &(\gamma_\mu{}^\nu)_{ij} P^\mu \wedge Q^i \wedge Q^j \otimes P_\nu\\
    &(\gamma^\mu)_{ij} P^\nu \wedge Q^i \wedge Q^j \otimes L_{\mu\nu}~.
  \end{aligned}
\end{equation}
The differential of the invariant subcomplex $(C^\bullet,d)$ is
defined by its action on the elements of $I^*$ and of $\fg$ as an
$I$-module; that is,
\begin{equation}
  \begin{aligned}[m]
    d P^\mu &= \half \gamma^\mu_{ij} Q^i \wedge Q^j\\
    d Q^i &= 0\\
    d P_\mu &= 0\\
    d Q_i &= - \gamma^\mu_{ij} Q^j \otimes  P_\mu\\
    d L_{\mu\nu} &= \eta_{\mu\rho} P^\rho \otimes P_\nu -
    \eta_{\nu\rho} P^\rho \otimes P_\mu + \half Q^i \otimes
    \gamma_{\mu\nu} \cdot Q_i\\
    d L_{ab} &= \half Q^i \otimes \gamma_{ab} \cdot Q_i~.
  \end{aligned}
\end{equation}

We now compute the differential $d:C^1 \to C^2$:
\begin{equation}
  \begin{aligned}[m]
    d(P^\mu \otimes P_\mu) &= \half \gamma^\mu_{ij} Q^i \wedge Q^j
    \otimes P_\mu\\
    d(Q^i \otimes Q_i) &= \gamma^\mu_{ij} Q^i \wedge Q^j \otimes P_\mu\\
    d(\varepsilon_\rho{}^{\mu\nu} P^\rho \otimes L_{\mu\nu}) &= - 2 \varepsilon_{\mu\nu}{}^\rho
    P^\mu \wedge P^\nu \otimes P_\rho - P^\mu
    \wedge Q^i \otimes \gamma_\mu \cdot Q_i - \half
    \gamma^{\mu\nu}_{ij} Q^i \wedge Q^j \otimes L_{\mu\nu}~.
  \end{aligned}
\end{equation}
We see that there is precisely one cocycle: $2 P^\mu \otimes P_\mu -
Q^i \otimes Q_i$, whence $H^1(\fg;\fg) \cong \RR$ in the absence of
any coboundaries.  As before, this outer derivation can be interpreted
as dilatations with the same weights as in the case of the Poincaré
superalgebra.  Since $\dim C^1 = 3$ and the space $Z^1$ of
$1$-cocycles has dimension $1$, we see that the space $B^2$ of
$2$-cocycles has dimension $2$.

To compute the differential $d:C^2 \to C^3$, we can recycle many of
the results from the similar calculation in §\ref{sec:minkowski}.
From the computation above of $d:C^1 \to C^2$, we learn that
$\gamma^\mu_{ij} Q^i \wedge Q^j \otimes P_\mu$ is a coboundary.  For
the remaining cochains one obtains
\begin{equation}
  \begin{aligned}[m]
    d(\varepsilon_{\mu\nu}{}^\rho P^\mu \wedge P^\nu \otimes P_\rho)
    &= - (\gamma_\mu{}^\nu)_{ij} P^\mu \wedge Q^i\wedge Q^j
    \otimes P_\nu~,\\
    d(P^\mu \wedge P^\nu \otimes L_{\mu\nu}) &= (\gamma^\mu)_{ij} Q^i
    \wedge Q^j \wedge P^\nu \otimes L_{\mu\nu} + \half P^\mu \wedge
    P^\nu \wedge Q^i \otimes \gamma_{\mu\nu} \cdot Q_i~,\\
    d(P^\mu \wedge Q^i \otimes \gamma_\mu \cdot Q_i) &=
    (\gamma_\mu{}^\nu)_{ij} P^\mu \wedge Q^i \wedge Q^j \otimes P_\nu
    + \half \gamma^\mu_{ij} Q^i \wedge Q^j \wedge Q^k \otimes
    \gamma_\mu
    \cdot Q_k~,\\
    d(\gamma^{\mu\nu}_{ij} Q^i \wedge Q^j \otimes L_{\mu\nu}) &= 2
    (\gamma_\mu{}^\nu)_{ij} P^\mu \wedge Q^i \wedge Q^j \otimes P_\nu
    + \half \gamma^{\mu\nu}_{ij} Q^i \wedge Q^j \wedge Q^k \otimes
    \gamma_{\mu\nu} \cdot Q_k~,\\
    d(\gamma^{ab}_{ij} Q^i \wedge Q^j \otimes L_{ab}) &= \half
    \gamma^{ab}_{ij} Q^i \wedge Q^j \wedge Q^k \otimes \gamma_{ab} \cdot Q_k\\
    &= -4 \gamma^\mu_{ij} Q^i \wedge Q^j \wedge Q^k \otimes \gamma_\mu
    \cdot Q_k~.
  \end{aligned}
\end{equation}

It is easy to construct a basis for the space $Z^2$ of cocycles:
\begin{equation}
  \begin{aligned}[m]
    &\gamma^\mu_{ij} Q^i \wedge Q^j \otimes P_\mu\\
    &\gamma^{\mu\nu}_{ij} Q^i \wedge Q^j \otimes L_{\mu\nu} + 2 P^\mu
    \wedge Q^i \otimes \gamma_\mu \cdot Q_i + 4
    \varepsilon_{\mu\nu}{}^\rho P^\mu \wedge P^\nu \otimes P_\rho\\
    &\gamma^{ab}_{ij} Q^i \wedge Q^j \otimes L_{ab} + 8 P^\mu \wedge
    Q^i \otimes \gamma_\mu \cdot Q_i + 8 \varepsilon_{\mu\nu}{}^\rho
    P^\mu \wedge P^\nu \otimes P_\rho~.
  \end{aligned}
\end{equation}
Thus $\dim Z^2 = 3$.  Since $\dim B^2 = 2$, with basis
\begin{equation}
  \begin{aligned}[m]
    &\gamma^\mu_{ij} Q^i \wedge Q^j \otimes P_\mu\\
    &2 \varepsilon_{\mu\nu}{}^\rho P^\mu \wedge P^\nu
    \otimes P_\rho - P^\mu \wedge Q^i \otimes \gamma_\mu \cdot Q_i +
    \half \gamma^{\mu\nu}_{ij} Q^i \wedge Q^j \otimes L_{\mu\nu}~,
  \end{aligned}
\end{equation}
we see that $\dim H^2(\fg;\fg) = 1$ and this allows us to conclude
that there is an infinitesimal deformation of the M2 Killing
superalgebra, with representative cocycle\footnote{There are other
  choices for representative cocycle, of course.  There is a choice
  where the $P\wedge P \otimes P$ term is absent and one might be
  puzzled at the fact that this seems to imply that there is no
  deformation to the bosonic subalgebra; however upon integrating that
  infinitesimal deformation, one is ineluctably led to adding those
  terms.}
\begin{equation}
      \gamma^{ab}_{ij} Q^i \wedge Q^j \otimes L_{ab} + 8 P^\mu \wedge
      Q^i \otimes \gamma_\mu \cdot Q_i + 8 \varepsilon_{\mu\nu}^\rho
      P^\mu \wedge P^\nu \otimes P_\rho~.
\end{equation}

In order to determine whether this deformation is integrable, let us
investigate the obstruction space $H^3(\fg;\fg)$.  Since $\dim Z^2 =
3$ and $\dim C^2 = 6$, it follows that $\dim B^3 = 3$.  As $\dim C^4
=4$, this means that $\dim H^3(I;\fg)^\fs \leq 1$ with equality if and
only if $d:C^3 \to C^4$ is the zero map.  A simple calculation shows
that, for instance,
\begin{multline}
  d (P^\mu \wedge P^\nu \wedge Q^i \otimes \gamma_{\mu\nu} \cdot Q_i)
  = \gamma^\mu_{ij} P^\nu \wedge Q^i \wedge Q^j \wedge Q^k \otimes
  \gamma_{\mu\nu} \cdot Q_k \\
  + 2 \gamma^\mu_{ij} P^\mu \wedge P^\nu \wedge Q^i \wedge Q^j \otimes
  P_\nu \neq 0~,
\end{multline}
whence $H^3(I;\fg)^\fs=0$ and the infinitesimal deformation is
unobstructed.  In fact, since $H^0(I;\fg)^\fs = 0$, we also see that
$H^3(\fg;\fg)=0$.

Integrating the infinitesimal deformation, we find the following
one-parameter ($t$) family of Lie superalgebras containing the M2
Killing superalgebra:
\begin{equation}
  \begin{aligned}[m]
    [Q_i, Q_j] &= \gamma^\mu_{ij} P_\mu - 2 t \gamma^{ab}_{ij} L_{ab}\\
    [P_\mu,P_\nu] &= 16 t \varepsilon_{\mu\nu}{}^\rho P_\rho\\
    [P_\mu,Q_i] &= - 8 t \gamma_\mu \cdot Q_i~,
  \end{aligned}
\end{equation}
where we have omitted the brackets involving the semisimple
generators, since these do not deform.  The parameter $t$ is mostly
fictitious: the resulting Lie algebras belong to three isomorphism
classes corresponding to $t>0$, $t=0$ and $t<0$.  Indeed, let us
rescale the generators $P_\mu \mapsto P'_\mu = \mu_P P_\mu$ and $Q_i
\mapsto Q'_i = \mu_Q Q_i$, while keeping $L'_{\mu\nu} = L_{\mu\nu}$
and $L'_{ab} = L_{ab}$ fixed.  Then choosing $\mu_P = -\frac1{8t}$
and $\mu_Q = \frac1{\sqrt{8|t|}}$ and dropping primes, we arrive
at the following normalised form for the superalgebra (for $t\neq 0$):
\begin{equation}
  \label{eq:m2deformed}
  \begin{aligned}[m]
    [Q_i, Q_j] &= \pm \left( \gamma^\mu_{ij} P_\mu + \tfrac14
      \gamma^{ab}_{ij} L_{ab} \right)\\
    [P_\mu,P_\nu] &= - 2 \varepsilon_{\mu\nu}{}^\rho P_\rho\\
    [P_\mu,Q_i] &= \gamma_\mu \cdot Q_i~,
  \end{aligned}
\end{equation}
where the sign is minus the sign of $t$.  The superalgebras for $t<0$
and $t>0$ are different real forms of the same complex Lie
superalgebra.  In fact, given any real Lie superalgebra, multiplying
the odd generators by $i$ gives another real Lie superalgebra,
reminiscent of the duality present in riemannian symmetric spaces.  We
notice that the Lie subalgebra spanned by $L_{\mu\nu}$ and $P_\mu$ is
isomorphic to $\fso(2,2)$.  This is easy to see as follows.  The
$P_\mu$ span a simple ideal isomorphic to $\fso(2,1)$ and the
$L_{\mu\nu}$ span a Lie algebra also isomorphic to $\fso(2,1)$.
Therefore their joint span is a semidirect product of $\fso(2,1)$ by
$\fso(2,1)$.  However, simple Lie algebras admit no outer derivations,
whence this semidirect product is actually isomorphic to a direct
product, whence $L_{\mu\nu}$ and $P_\mu$ span a Lie subalgebra
isomorphic to $\fso(2,1) \oplus \fso(2,1) \cong \fso(2,2)$.  This
isomorphism can be made manifest by noticing that $P_\mu$ and $R_\mu
:= P_\mu - \varepsilon_\mu{}^{\nu\rho} L_{\nu\rho}$ are commuting
$\fso(2,1)$-subalgebras.  In particular, $R_\mu$ acts trivially on the
supercharges.  The Lie superalgebra spanned by $P_\mu$, $Q_\alpha$ and
$L_{ab}$ is isomorphic to a real form of the the classical Lie
superalgebra $D(4,1)$ in \cite{KacSuperSketch}.  Hence abstractly as a
Lie superalgebra the deformed M2-brane superalgebra is isomorphic to a
real form of $A_1 \oplus D(4,1)$.  The $\fso(2,2)$ subalgebra of
the deformed superalgebra suggests that quantum corrections curve
the brane worldvolume to $\AdS_3$ with the quantum parameter being
related to the curvature of the $\AdS_3$.  Another possibility,
currently being investigated \cite{FigRitDef}, would be that this
deformation is the Killing superalgebra of a one-parameter family of
\emph{classical} half-BPS M2-brane backgrounds where the M2-brane
wraps an $\AdS_3$.

\subsection{Rigidity of the M5-brane Killing superalgebra}
\label{sec:M5}

The Killing superalgebra $\fg = \fg_0 \oplus \fg_1$ of the M5-brane is
the subalgebra of the Poincaré superalgebra defined as follows.  First
we split the $11$-dimensional lorentzian vector space $V = W \oplus
W^\perp$, where $W$ is six-dimensional lorentzian.  The subalgebra of
the Lorentz algebra which preserves this split is $\fso(W) \oplus
\fso(W^\perp)$.  Then $\fg_0 = \fso(W) \oplus \fso(W^\perp) \oplus W$.
The odd part of the superalgebra is the subspace $\Delta$ of the
$\fso(V)$ spinor module $\Delta(V)$ consisting of those spinors $\psi$
for which $\bnu_W \cdot \psi = \psi$.  The volume element $\bnu_W$ is
skew-adjoint with respect to the invariant symplectic form and
satisfies $\bnu_W^2 = 1$.  As an $\fso(W) \oplus
\fso(W^\perp)$-module, this is $[\Delta(W)_+ \otimes
\Delta(W^\perp)]$, where the chirality condition on the $\fso(W)$
spinors is precisely the supersymmetry projection condition
$\bnu_W\cdot \psi = \psi$, and where the brackets denote the
underlying real representation of the product of quaternionic
representations $\Delta(W)_+$ and $\Delta(W^\perp)$, each of which has
four complex dimensions.  The resulting superalgebra has dimension
$(31|16)$.

As before, we let $I$ be the supertranslation ideal isomorphic to $W
\oplus \Delta$, so that again $\fs = \fso(W)\oplus \fso(W^\perp)$.  We
will let $P_\mu$ and $Q_i$ denote a basis for $I$ and $L_{\mu\nu}$ and
$L_{ab}$ be a basis for $\fs$.  As before we will let $P^\mu$ and
$Q^i$ denote the bases for $I^*$ canonically dual to $P_\mu$ and
$Q_i$, respectively.

In this basis, the Lie brackets are formally the same as those in
\eqref{eq:SLAM2} after suitably reinterpreting the symbols.

There are no $\fs$-invariant elements in $\fg$, whence there are no
invariant cochains.  The space $C^1$ of invariant $1$-cochains is
two-dimensional, spanned by the identity maps $W \to W$ and $\Delta
\to \Delta$.  The representative cochains are
\begin{equation}
  P^\mu \otimes P_\mu \qquad\text{and}\qquad Q^i \otimes Q_i~.
\end{equation}

The space $C^2$ of invariant $2$-cochains is three-dimensional,
spanned by the identity map $\Lambda^2W \to \Lambda^2W$, the Clifford
product $W \otimes \Delta \to \Delta$, and the squaring map $S^2\Delta
\to W$, with representative cochains
\begin{equation}
  P^\mu \wedge P^\nu \otimes L_{\mu\nu}~,\quad
  P^\mu \wedge Q^i \otimes \gamma_\mu \cdot Q_i\quad\text{and}\quad
  \gamma^\mu_{ij} Q^i \wedge Q^j \otimes P_\mu~.
\end{equation}
The squaring map $S^2\Delta \to \Lambda^2 W \oplus \Lambda^2 W^\perp$ is zero
because of the projection condition on the spinors.

The calculation of the differential on $C^1$ and $C^2$ is very similar
to those for the M2 brane and can almost be read off from those.
There is a $1$-cocycle
\begin{equation}
  2 P^\mu \otimes P_\mu - Q^i \otimes Q_i~,
\end{equation}
whence $H^1(\fg;\fg) \cong \RR$ in the absence of any coboundaries.
As before, this outer derivation can be interpreted as dilatations
with the same weights as in the case of the Minkowski and
the M2-brane Killing superalgebras.

We learn that $\gamma^\mu_{ij} Q^i \wedge Q^j \otimes P_\mu$ is the
only $2$-coboundary, whereas the calculations
\begin{equation}
  d(P^\mu \wedge P^\nu \otimes L_{\mu\nu}) = \gamma^\mu_{ij} Q^i
  \wedge Q^j \wedge P^\nu \otimes L_{\mu\nu} + \half P^\mu \wedge
  P^\nu \wedge Q^i \otimes \gamma_{\mu\nu} \cdot Q_i
\end{equation}
and
\begin{equation}
  d(P^\mu \wedge Q^i \otimes \gamma_\mu \cdot Q_i) =  \half
  \gamma^\mu_{ij} Q^i \wedge Q^j \wedge Q^k \otimes \gamma_\mu \cdot
  Q_k
\end{equation}
show that there are no further $2$-cocycles.  Therefore
$H^2(\fg;\fg)=0$ and the M5 brane Killing superalgebra is rigid.

\section{Superalgebra deformations of purely gravitational
  backgrounds}
\label{sec:KSAgrav}

In this section we tackle the Lie superalgebra deformations of the
purely gravitational $\half$-BPS backgrounds: the Kaluza--Klein
monopole \cite{SMKK,GPMKK,HKMKK} and the M-wave \cite{Mwave}.   In the
absence of flux, the Killing spinors are parallel in these
backgrounds.  This means that the Lie bracket of supercharges consists
of parallel vectors and hence of translations.

\subsection{A deformation of the M-wave Killing superalgebra}
\label{sec:MWave}

The Killing superalgebra $\fg=\fg_0 \oplus \fg_1$ of the maximally
symmetric $\half$-BPS M-wave is the $(38|16)$-dimensional
subsuperalgebra of the Poincaré superalgebra defined as follows.  We
first split the eleven-dimensional lorentzian vector space $V = W
\oplus W^\perp$, where $W$ is a two-dimensional lorentzian subspace
and $W^\perp$ is the perpendicular euclidean space, which can be
interpreted as the transverse space to the wave front.  We can write
$W = W_+ \oplus W_-$, where $W_\pm$ are isotropic one-dimensional
subspaces, with $W_+$ spanned by the parallel vector.  The even
subalgebra $\fg_0 = \fso(W^\perp) \oplus W$ and the odd subspace
$\fg_1=\Delta$, with $\Delta$ the sixteen-dimensional subspace of the
space of eleven-dimensional spinors defined as the kernel of Clifford
multiplication by $W_+$.  As before, we take $I \cong W \oplus \Delta$
to be the supertranslation ideal and $\fs = \fso(W^\perp)$ to be the
semisimple factor.  Let $\be_\pm$ span $W_\pm$, $\be_a$ span $W$ and
$\beps_i$ span $\Delta$.  The corresponding basis for $I$ is given by
$P_\pm$ and $Q_i$, with $P^\pm$ and $Q^i$ denoting the canonical dual
basis for $I^*$.  We will let $L_{ab}$ span $\fso(W^\perp)$.

In this basis, the Lie brackets take the form
\begin{equation}
  [L_{ab} , Q_i] = \half \gamma_{ab}\cdot Q_i
  \qquad\text{and}\qquad
  [Q_i, Q_j] = \Omega_{ij} P_+~,
\end{equation}
in addition to the ones of $\fs$, where the bilinear form $\Omega_{ij}
:= \left<\beps_i,\gamma^+ \cdot \beps_j\right> =
\left<\beps_i,\gamma_- \cdot \beps_j\right>$ is symmetric and
positive-definite on $\Delta$.  As representations of $\fso(W^\perp)$,
we have
\begin{equation}
  S^2\Delta \cong \RR \oplus W^\perp \oplus \Lambda^4 W^\perp~.
\end{equation}

The differential of the invariant subcomplex $(C^\bullet,d)$ is
defined by its action on the elements of $I^*$ and of $\fg$ as an 
$I$-module; that is,
\begin{equation}
  \begin{aligned}[m]
    d P^- &= 0\\
    d Q^i &= 0\\
    d P_\pm &= 0
  \end{aligned}
  \qquad\qquad
  \begin{aligned}[m]
    d P^+ &= \half \Omega_{ij} Q^i \wedge Q^j\\
    d Q_i &= - \Omega_{ij} Q^j \otimes  P_+ \\
    d L_{ab} &= \half Q^i \otimes \gamma_{ab} \cdot Q_i~.
  \end{aligned}
\end{equation}

The space of invariant $0$-cochains is two-dimensional, spanned by
$P_\pm$.  They are central elements in $\fg$, whence cocycles.  Since
there are no coboundaries, $\dim H^0(\fg;\fg) = 2$ and $\dim B^1=0$.

The space of invariant $1$-cochains is $5$-dimensional, consisting of
the $4$-dimensional subspace $\End(W)$ and the one-dimensional
subspace spanned by the identity map $\Delta \to \Delta$.  The
corresponding cochains are $Q^i \otimes Q_i$ and $P^\pm\otimes P_\pm$
with uncorrelated signs.  The differential $d:C^1 \to C^2$ is such
that $P^- \otimes P_\pm$ are cocycles and
\begin{equation}
  d(P^+\otimes P_\pm) = \half \Omega_{ij} Q^i \wedge Q^j \otimes P_\pm~,
\end{equation}
and
\begin{equation}
  d(Q^i \otimes Q_i) = \Omega_{ij} Q^i \wedge Q^j \otimes P_+~.
\end{equation}
Therefore we see that $\dim Z^1 = 3$, whence $\dim B^2 = 2$, with
basis $P^- \otimes P_\pm$ and
\begin{equation}
  Q^i \otimes Q_i + 2 P^+ \otimes P_+~.
\end{equation}
As there are no coboundaries, we see that $\dim H^1(\fg;\fg) = 3$.

The space of invariant $2$-cochains is six-dimensional with basis
\begin{equation}
  P^+\wedge P^- \otimes P_\pm \qquad
  P^\pm \wedge Q^i \otimes Q_i \qquad
  \Omega_{ij}  Q^i \wedge Q^j \otimes P_\pm~.
\end{equation}
The differential $d:C^2 \to C^3$ is given by
\begin{equation}
  \begin{aligned}[m]
    d \left(P^+ \wedge P^- \otimes P_\pm\right) &= \half \Omega_{ij}
    P^-
    \wedge Q^i \wedge Q^j  \otimes P_\pm\\
    d \left(P^- \wedge Q^i \otimes Q_i\right) &= - \Omega_{ij} P^-
    \wedge Q^i \wedge Q^j  \otimes P_+\\
    d \left(P^+ \wedge Q^i \otimes Q_i\right) &= - \Omega_{ij} P^+
    \wedge Q^i \wedge Q^j \otimes P_+ + \half \Omega_{ij} Q^i \wedge
    Q^j \wedge Q^k \otimes Q_k~,
  \end{aligned}
\end{equation}
and by $d\left( \Omega_{ij}  Q^i \wedge Q^j \otimes P_\pm\right) = 0$.
These two cocycles are also coboundaries and the only other cocycle is
\begin{equation}
  P^-\wedge Q^i \otimes Q_i + 2 P^+ \wedge P^- \otimes P_+~.
\end{equation}
In other words, $\dim Z^2 = 3$ and, since $\dim B^2 = 2$, we see that
$\dim H^2(\fg;\fg) = 1$ with the above representative cocycle.  This
means that there is a one-dimensional space of infinitesimal
deformations.  It is easy to show by an explicit computation that this
infinitesimal deformation is unobstructed and we end up with the
following one-parameter ($t$) family of Lie superalgebras containing
the M-wave Killing superalgebra:
\begin{equation}
  [Q_i, Q_j] = \Omega_{ij} P_+\qquad
  [P_-, Q_i] = - t Q_i\qquad
  [P_+, P_-] = 2 t P_+~,
\end{equation}
where we have omitted the brackets involving $\fs$, as these remain
undeformed.  By rescaling $P_-$ we see that all superalgebras for $t
\neq 0$ are isomorphic, whence we can let $t$ above take only two
values: $0$ and $1$.  In the former case, it is the original M-wave
Killing superalgebra, whereas in the latter case it is a deformation
\begin{equation}
  \label{eq:MWdeformed}
  \begin{aligned}[m]
    [Q_i, Q_j] &= \Omega_{ij} P_+\\
    [P_-, Q_i] &= Q_i\\
    [P_-, P_+] &= 2 P_+~,
  \end{aligned}
\end{equation}
perhaps induced by quantum corrections or perhaps belonging to a
one-parameter family of backgrounds which tends to the M-wave under
some geometric limit contracting its Killing superalgebra.

\subsection{Deformations of the Kaluza--Klein monopole Killing superalgebra}
\label{sec:KKmonopole}

The Killing superalgebra $\fg=\fg_0 \oplus \fg_1$ of the $\half$-BPS
Kaluza--Klein monopole is the $(32|16)$-dimensional subsuperalgebra of
the Poincaré superalgebra defined as follows.  Let us split the
$11$-dimensional lorentzian vector space $V = W \oplus W^\perp$, where
$W$ is a $7$-dimensional lorentzian subspace and $W^\perp$ is the
perpendicular $4$-dimensional euclidean space.  The even subalgebra
$\fg_0 = \fso(W) \oplus W \oplus \fu(W^\perp)$, where
$\fu(W^\perp)\subset \fso(W^\perp)$ is the $4$-dimensional subalgebra
preserving a self-dual hermitian structure on $W^\perp$.  If the
hermitian structure is defined by the metric and a compatible complex
structure $J$, then the associated $2$-form $\omega$ on $W^\perp$ is
anti-self dual: $\omega \in \Lambda_-^2W^\perp$.  Then the subalgebra
$\fu(W^\perp)$ is spanned by the self-dual two forms together with
$\omega$.  Under this decomposition, we will write $\fu(W^\perp) =
\fsu(W^\perp) \oplus \RR\omega$.

The odd subspace is $\fg_1 \cong \Delta$, with $\Delta$ the
sixteen-dimensional subspace of the space of eleven-dimensional
spinors defined by the projection condition $\bnu_{W^\perp} \cdot
\psi= - \psi$.  We take $I \cong W \oplus \Delta \oplus \RR\omega$ to
be the ideal and $\fs = \fso(W) \oplus \fsu(W^\perp)$ to be the
semisimple factor.  We will let $P_\mu$, $Q_i$ and $\omega$ span $I$
and $L_{\mu\nu} := L(\be_\mu \wedge \be_\nu)$, for $\mu<\nu$, and
$L_{ab}^+ := L\left(\be_a \wedge \be_b + \star (\be_a \wedge
  \be_b)\right)$, for $a<b$, span $\fs$.  We let $P^\mu$, $Q^i$ and
$\omega^*$ denote the canonical dual basis for $I^*$.

In this basis, the Lie brackets are given by
\begin{equation}
  \label{eq:SLAMKK}
  \begin{aligned}[m]
    [L_{\mu\nu} , Q_i] &= \half \gamma_{\mu\nu}\cdot Q_i\\
    [L^+_{ab} , Q_i] &= \half \gamma^+_{ab}\cdot Q_i
  \end{aligned}
  \qquad\qquad
  \begin{aligned}[m]
    [L_{\mu\nu}, P_\rho] &= \eta_{\nu\rho} P_\mu - \eta_{\mu\rho} P_\nu\\
    [Q_i, Q_j] &= \gamma^\mu_{ij} P_\mu
  \end{aligned}
\end{equation}
in addition to the ones of $\fs$.

As an $\fs$-module, $\Delta = [\Delta^{1,6} \otimes \Delta^4_-]$, with
$\Delta^{1,6}$ the complex $8$-dimensional quaternionic spinorial
representation of $\Spin(1,6)$, $\Delta^4_-$ the complex
$2$-dimensional quaternionic representation of $\Spin(4)$ consisting
of negative chirality spinors, and where as usual the brackets
indicate the underlying real subrepresentation.  As an $\fs$-module,
\begin{equation}
  S^2\Delta \cong W \oplus \Lambda^2W \oplus \Lambda^2_+W^\perp \oplus
  \left(\Lambda^3W \otimes \Lambda^2_+W^\perp\right)~.
\end{equation}

It is now possible to list the $\fs$-invariant cochains.  The centre
of $\fg$ is spanned by $\omega$, whence $\dim C^0 = \dim Z^0 = 1$ and
in the absence of coboundaries $\dim H^0(\fg;\fg) = 1$.  This also
shows that $\dim B^1 = 0$.  The invariant $1$-cochains are induced by
the identity maps $W \to W$, $\Delta \to \Delta$ and $\RR\omega \to
\RR\omega$, yielding the following cochains
\begin{equation}
  P^\mu \otimes P_\mu \qquad
  Q^i \otimes Q_i \qquad
  \omega^* \otimes \omega~,
\end{equation}
whence $\dim C^1 = 3$.

The invariant $2$-cochains are given by the
natural isomorphism $\Lambda^2 W \to \fso(W)$, Clifford multiplication
$W \otimes \Delta \to \Delta$, the squaring maps $S^2 \Delta \to W
\oplus \fso(W) \oplus \fsu(W^\perp)$, as well as the isomorphisms $\RR\omega
\otimes W \to W$ and $\RR\omega \otimes \Delta \to \Delta$ induced by
the identity maps on $W$ and $\Delta$.  The corresponding cochains are
\begin{equation}
  \begin{aligned}[m]
    \gamma^\mu_{ij} Q^i \wedge Q^j \otimes P_\mu\\
    \gamma^{\mu\nu}_{ij} Q^i \wedge Q^j \otimes L_{\mu\nu}\\
    \left(\gamma^{ab}_+\right)_{ij} Q^i \wedge Q^j \otimes L_{ab}^+\\
  \end{aligned}
  \qquad\qquad
  \begin{aligned}[m]
    P^\mu \wedge P^\nu \otimes L_{\mu\nu}\\
    P^\mu \wedge Q^i\otimes \gamma_\mu \cdot Q_i\\
    \omega^* \wedge P^\mu \otimes P_\mu\\
    \omega^* \wedge Q^i \otimes Q_i~,
  \end{aligned}
\end{equation}
whence $\dim C^2 = 7$.

The space of invariant $3$-cochains is $9$-dimensional, spanned by
the following cochains associated to the natural maps:
\begin{equation}
  \begin{aligned}[m]
    \omega^* \wedge P^\mu \wedge Q^i \otimes \gamma_\mu \cdot Q_i\\
    \omega^* \wedge P^\mu \wedge P^\nu \otimes L_{\mu\nu}\\
    \gamma^\mu_{ij}\omega^* \wedge Q^i \wedge Q^j \otimes P_\mu\\
    \gamma^{\mu\nu}_{ij} \omega^* \wedge Q^i \wedge Q^j \otimes
    L_{\mu\nu}\\
    \left(\gamma^{ab}_+\right)_{ij} \omega^* \wedge Q^i \wedge Q^j
    \otimes L_{ab}^+
  \end{aligned}
  \qquad\qquad
  \begin{aligned}[m]
    P^\mu \wedge P^\nu \wedge Q^i \otimes \gamma_{\mu\nu}\cdot Q_i\\
    \gamma^\mu_{ij} P^\nu \wedge Q^i \wedge Q^j \otimes L_{\mu\nu}\\
    \left(\gamma_\mu^\nu\right)_{ij} P^\mu \wedge Q^i \wedge Q^j
    \otimes P_\nu\\
    \gamma^\mu_{ij} Q^i \wedge Q^j \wedge Q^k \otimes \gamma_\mu \cdot
    Q_k~,
  \end{aligned}
\end{equation}
where the absence of the cochain
\begin{equation}
  \left(\gamma^{ab}_+\right)_{ij} Q^i \wedge Q^j \wedge Q^k \otimes
  \gamma^+_{ab} \cdot Q_k
\end{equation}
is explained by the fact that it is in the span of the above cochains
by virtue of the Fierz identity \eqref{eq:fierz}, and the absence of
the cochain
\begin{equation}
    \gamma^{\mu\nu}_{ij} Q^i \wedge Q^j \wedge Q^k \otimes
    \gamma_{\mu\nu} \cdot Q_k
\end{equation}
is explained by the Fierz identity
\begin{equation}
  \label{eq:fierzMKK1}
   \half \left<\psi, \gamma^{\mu\nu} \psi \right> \gamma_{\mu\nu} \psi
   =  -3 \left<\psi, \gamma^\mu \psi \right> \gamma_\mu \psi~,
\end{equation}
for $\psi \in \Delta$.  Together with the Fierz identity
\eqref{eq:fierz}, we also obtain
\begin{equation}
  \label{eq:fierzMKK2}
   \half \left<\psi, \gamma_+^{ab} \psi \right> \gamma^+_{ab} \psi
   =  -2 \left<\psi, \gamma^\mu \psi \right> \gamma_\mu \psi~.
\end{equation}

The differential of the invariant subcomplex $(C^\bullet,d)$ is
defined by its action on the elements of $I^*$ and of $\fg$ as an 
$I$-module; that is,
\begin{equation}
  \begin{aligned}[m]
    d P_\mu &= 0\\
    d Q^i &= 0\\
    d \omega^* &= 0\\
    d \omega &=0
  \end{aligned}
  \qquad\qquad
  \begin{aligned}[m]
    d P^\mu &= \half \gamma^\mu_{ij} Q^i \wedge Q^j\\
    d Q_i &= - \gamma^\mu_{ij} Q^j \otimes  P_\mu \\
    d L_{\mu\nu} &= \eta_{\mu\rho} P^\rho \otimes P_\nu -
    \eta_{\nu\rho} P^\rho \otimes P_\mu + \half Q^i \otimes
    \gamma_{\mu\nu} \cdot Q_i\\
    d L^+_{ab} &= \half Q^i \otimes \gamma^+_{ab} \cdot Q_i~.
  \end{aligned}
\end{equation}

The unique invariant $0$-cochain $\omega$ is a cocycle, whence $\dim
H^0(\fg;\fg) = 1$.  The differential $d:C^1 \to C^2$ is given by
\begin{equation}
  \begin{aligned}[m]
    d (P^\mu \otimes P_\mu) &= -\half \gamma^\mu_{ij} Q^i \wedge Q^j
    \otimes P_\mu\\
    d ( Q^i \otimes Q_i) &= \gamma^\mu_{ij} Q^i \wedge Q^j \otimes
    P_\mu\\
    d ( \omega^* \otimes \omega) &= 0~.
  \end{aligned}
\end{equation}
The space of cocycles is $2$-dimensional, spanned by $\omega^* \otimes
\omega$ and
\begin{equation}
  2 P^\mu \otimes P_\mu - Q^i \otimes Q_i~.
\end{equation}
Since there are no coboundaries, $\dim H^1(\fg;\fg) = 2$.  This
calculation also shows that $\dim B^2 = 1$, spanned by
$\gamma^\mu_{ij} Q^i \wedge Q^j \otimes P_\mu$.

The differential $d:C^2 \to C^3$ on the remaining cochains is given by
\begin{equation}
  \begin{aligned}[m]
    d(P^\mu \wedge P^\nu \otimes L_{\mu\nu}) &= \gamma^\mu_{ij} Q^i
    \wedge Q^j \wedge P^\nu \otimes L_{\mu\nu} + \half P^\mu \wedge
    P^\nu \wedge Q^i \otimes \gamma_{\mu\nu} \cdot Q_i\\
    d (\omega^* \wedge P^\mu \otimes P_\mu) &= - \half \gamma^\mu_{ij}
    \omega^* \wedge Q^i \wedge Q^j \otimes P_\mu\\
    d (\omega^* \wedge Q^i \otimes Q_i) &= - \gamma^\mu_{ij}\omega^*
    \wedge Q^i \wedge Q^j \otimes P_\mu\\
    d(P^\mu \wedge Q^i \otimes \gamma_\mu \cdot Q_i) &=
    \left(\gamma_\mu{}^\nu\right)_{ij} P^\mu \wedge Q^i \wedge Q^j
    \otimes P_\nu + \half \gamma^\mu_{ij} Q^i \wedge Q^j \wedge Q^k
    \otimes \gamma_\mu \cdot Q_k\\
    d(\gamma^{\mu\nu}_{ij} Q^i \wedge Q^j \otimes L_{\mu\nu}) &= 2
    \left(\gamma_\mu{}^\nu\right)_{ij} P^\mu \wedge Q^i \wedge Q^j
    \otimes P_\nu -3 \gamma^\mu_{ij} Q^i \wedge Q^j \wedge Q^k \otimes
    \gamma_\mu \cdot Q_k\\
    d\left(\left(\gamma_+^{ab}\right)_{ij} Q^i \wedge Q^j \otimes
      L^+_{ab}\right) &= - 2 \gamma^\mu_{ij} Q^i \wedge Q^j \wedge Q^k
    \otimes \gamma_\mu \cdot Q_k~,
  \end{aligned}
\end{equation}
where we have used equations \eqref{eq:fierzMKK1} and
\eqref{eq:fierzMKK2}.  It is not hard to show that there are two
linearly independent cocycles:
\begin{equation}
  \label{eq:MKKcoc1}
  2 \omega^* \wedge P^\mu \otimes P_\mu - \omega^* \wedge Q^i \otimes
  Q_i~,
\end{equation}
and
\begin{equation}
  \label{eq:MKKcoc2}
  P^\mu \wedge Q^i \otimes \gamma_\mu\cdot Q_i - \half
  \gamma^{\mu\nu}_{ij} Q^i \wedge Q^j \otimes L_{\mu\nu} + 
  \left(\gamma_+^{ab}\right)_{ij} Q^i \wedge Q^j \otimes L^+_{ab}~,
\end{equation}
whence $\dim H^2(\fg;\fg) = 2$.  This gives rise to a two-dimensional
space of infinitesimal deformations of the Killing superalgebra.

The deformation corresponding to the cocycle \eqref{eq:MKKcoc1} is
unobstructed, which gives rise to a one-parameter $(t)$ deformation of
the Killing superalgebra of the Kaluza--Klein monopole, given by
\begin{equation}
  [Q_i, Q_j] = \gamma^\mu_{ij} P_\mu \qquad
  [\omega, Q_i] = t Q_i\qquad
  [\omega, P_\mu] = 2 t P_\mu~,
\end{equation}
together with the brackets involving $\fs$, which do not deform.  By
rescaling $\omega$, we see that there are two isomorphism classes of
Lie superalgebras in this family, corresponding to the values $t=0$,
which is the original Killing superalgebra, and $t=1$, given by
\begin{equation}
  \label{eq:MKKdeform1}
  \begin{aligned}[m]
  [Q_i, Q_j] &= \gamma^\mu_{ij} P_\mu \\
  [\omega, Q_i] &= Q_i\\
  [\omega, P_\mu] &= 2 P_\mu~,
  \end{aligned}
\end{equation}
in addition to the brackets involving $\fs$.  In essence $\omega$ acts
now as homotheties on the $7$-dimensional Minkowski spacetime as well
as rotations in Taub-NUT.  Curiously it is now seen to generate a
subgroup $\RR$ and not a compact subgroup $\U(1)$.   The form of the
superalgebra would suggest a geometry which is no longer a metric
product but rather a \emph{warped} product where the size of the
Minkowski factor now depends on the angular variable in the Taub-NUT;
however this cannot be the case because the identifications in the
angular variable.  This is reminiscent of a non-geometric background
\cite{AtishChrisTNG} and might be related to the discussion in
\cite{TongTNS5,GHMTKKM}.

The deformation corresponding to the cocycle \eqref{eq:MKKcoc2} is
also unobstructed, but unlike the previous case, this requires adding
a term of order $t^2$ to the $[P,P]$ bracket.  The one-parameter
deformation is given by
\begin{equation}
  \begin{aligned}[m]
  [Q_i, Q_j] &= \gamma^\mu_{ij} P_\mu + t \gamma^{\mu\nu}_{ij}
  L_{\mu\nu} - 2t \left(\gamma^{ab}_+\right)_{ij} L_{ab}^+\\
  [P_\mu, Q_i] &= -t \gamma_\mu \cdot Q_i\\
  [P_\mu, P_\nu] &= 4 t^2 L_{\mu\nu}~.
  \end{aligned}
\end{equation}
For $t\neq0$, we may rescaling $P_\mu$ by $\frac1{2t}$ and $Q_i$ by
$\frac1{\sqrt{2|t|}}$ in order to bring the Lie algebra to the
following form
\begin{equation}
  \label{eq:MKKdeform2}
  \begin{aligned}[m]
  [Q_i, Q_j] &= \pm \left( \gamma^\mu_{ij} P_\mu + \half \gamma^{\mu\nu}_{ij}
  L_{\mu\nu} - \left(\gamma^{ab}_+\right)_{ij} L_{ab}^+ \right)\\
  [P_\mu, Q_i] &= -\half \gamma_\mu \cdot Q_i\\
  [P_\mu, P_\nu] &= L_{\mu\nu}~,
  \end{aligned}
\end{equation}
where the sign is the sign of $t$.  These Lie superalgebras are real
forms of the classical Lie superalgebra $D(4,1)$ augmented by the
central element $\omega$.  In particular the even subalgebra is
$\fso(2,6) \oplus \fu(2)$, which suggests that the Minkowski factor
deforms to $\AdS_7$.

Any other linear combination of the cocycles \eqref{eq:MKKcoc1} and
\eqref{eq:MKKcoc2} is obstructed.  This is easy to see because the
weights of $Q$ and $P$ relative to the adjoint action of $\omega$
implied by the cocycle \eqref{eq:MKKcoc1} is incompatible with the
$[P,Q]$ bracket implied by the cocycle \eqref{eq:MKKcoc2}.

\section{Conclusion}
\label{sec:conclusion}

In this paper we have started the study of the deformations of Killing
superalgebras of supersymmetric eleven-dimensional backgrounds.  Our
motivation is that the deformed Killing superalgebra gives us hints
about the possible quantum corrections a classical background might
undergo or, keeping within the classical theory, about possible
backgrounds which can tend to the original one under a geometric
limit---the rationale being that the Killing superalgebra contracts
under such a limit whence it must be found among the deformations of
the contracted algebra.

We have studied backgrounds with either maximal supersymmetry or
half-BPS.  We have shown the rigidity of the Killing superalgebras of
the maximally supersymmetric backgrounds we have studied, namely
Minkowski spacetime and the Freund--Rubin backgrounds.  This agrees
with the heuristic idea that supersymmetry tends to rigidify the
geometry and with known results about the fact that these backgrounds
do not admit quantum corrections \cite{KalRajVacua}.  We have not
attempted to classify the deformations of the Killing superalgebra of
the Kowalski-Glikman background, since the semisimple factor is not
large enough to allow a painless calculation, but since it is know
that its superalgebra gets deformed \cite{ShortLimits, Limits,
  HatKamiSaka,   FSPL}, we know that it will have at least two
deformations: corresponding to the Killing superalgebras of the
Freund--Rubin backgrounds.  However I would hazard the ``conjecture''
that no further deformations exist.

Among the half-BPS backgrounds considered in this paper, the Killing
superalgebra of the M5-brane is rigid, whereas that of the M2-brane,
the M-wave and the Kaluza--Klein monopole admits nontrivial
deformations given up to isomorphism, by the Lie superalgebras in
equations (\ref{eq:m2deformed}), (\ref{eq:MWdeformed}),
(\ref{eq:MKKdeform1}) and (\ref{eq:MKKdeform2}).  In particular, the
structure of the Lie superalgebras in (\ref{eq:m2deformed}) and
(\ref{eq:MKKdeform2}) suggest that the worldvolume of the M2-brane
acquires constant negative curvature and so does that of the Minkowski
factor in the Kaluza--Klein monopole.  In a forthcoming paper
\cite{FigVer10d}, in which we study the Killing superalgebra
deformations of some ten-dimensional supergravity backgrounds, we
present evidence that deformations behave well under Kaluza--Klein
reduction, suggesting a geometric interpretation for the deformed
superalgebras found here and also in that paper.  We are currently
investigating whether the existence of the deformations found in this
paper can be explained within the context of supergravity
\cite{FigRitDef}.

\section*{Acknowledgments}

I have benefited from discussions with Chris Hull, Hyakutake
Yoshifumi, Patricia Ritter, Joan Simón and Bert Vercnocke.

\bibliographystyle{utphys}
\bibliography{AdS,AdS3,ESYM,Sugra,Geometry,Algebra}

\end{document}